\documentclass[%
    aip,
    jcp,
    amsmath,
    amssymb,
    twocolumn,
    superscriptaddress,
    10pt]{revtex4-1}
\usepackage{dcolumn}
\usepackage{graphicx} 
\usepackage{bm}
\usepackage{longtable}
\usepackage[english]{babel}

\begin{document}

\title{Terahertz absorption of lysozyme in solution} 

\author{Daniel R.\ Martin}
\affiliation{Department of Physics, Arizona State University, PO Box 871504, Tempe, Arizona 85287}
\author{Dmitry V.\ Matyushov}
\affiliation{Department of Physics and School of Molecular Sciences, 
         Arizona State University, PO Box 871504, Tempe, Arizona 85287}
\email{dmitrym@asu.edu}
\begin{abstract}
Absorption of radiation by solution is described by the solution dielectric constant and can be viewed as a specific application of the dielectric theory of solutions. For ideal solutions, the dielectric boundary value problem separates the polar response into the polarization of the void in the liquid created by the solute and the response of the solute dipole. In the case of a  protein as a solute, its nuclear dynamics do not project on significant fluctuations of the dipole moment in the terahertz domain of frequencies and the protein dipole can be viewed as dynamically frozen. Absorption of radiation then reflects the interfacial polarization. Here we apply an analytical theory and computer simulation to absorption of radiation by ideal solutions of lysozyme. Comparison with experiment shows that Maxwell electrostatics fails to describe the polarization of the protein-water interface and the ``Lorentz void'', which does not anticipate polarization of the solute void by the external field (no surface charges), better represents the data. An analytical theory for the slope of the solution absorption against the volume fraction of the solute is formulated in terms of the cavity field response function. It is calculated from molecular dynamics simulations in good agreement with experiment. The protein hydration shell emerges as a separate sub-ensemble, which collectively is not described by the standard electrostatics of dielectrics.      
\end{abstract}

\maketitle

\section{Introduction}
\label{sec:1}
Inserting a solute into a molecular liquid leads to a nonlinear perturbation of its density.\cite{Chandler:93} The density profile formed around a solute is a sensitive function of the strength of the intermolecular potential. Weak dewetting\cite{ChandlerNature:05,GardePRL:09} or, alternatively, collapse of hydration shells\cite{Brovchenko:2007jc,BrovchenkoBook,DMjpcl2:12} occurs for hydration when the solute-water interaction is either weaker (dewetting) or stronger (collapse) than the water-water interaction. 

In contrast to the diversity of scenarios for the density profile of hydration layers, the orientational structure of the solute-solvent interface is typically viewed as a specific case of the standard dielectric boundary-value problem. The dipoles of water then orient along the lines of the solute's electric field according to the rules of electrostatics of dielectrics.\cite{Jackson:99} There are a number of fundamental reasons why this simplistic picture can break down,\cite{DMjcp3:14} particularly in the case of patterned interfaces of nano-scale dimension.\cite{Giovambattista:08} By the patterned interface we mean a substrate with its polarity or charge changing on the length-scale comparable to the correlation length between the dipoles in the surrounding water. The most illuminating example is a patchwork of positive and negative charges, each orienting the nearest water molecules along the field of the surface charge. If the distance between different charges is comparable with the size of the surface polarized domain, this spatial arrangement will lead to the orientational frustration of the water molecules at the domain boundaries.  One can anticipate that the response of this patchwork of frustrated domains can potentially be different from the response of a homogeneous dielectric. 

The picture of a patchwork of positive and negative charges distributed over a nanometer-scale surface is a good match of the protein-water interface. Proteins possess a large density of surface ionized residues, which provides the free energy required to stabilize a protein in solution. The distribution of positive and negative charges at the surface of a globular protein is nearly uniform, with the average distance between the charges $\sim 10$ \AA.\cite{Barlow:87,Takashima:2002ka} This distance is comparable to the length of a chain of 3--4 water molecules. When the two waters at the ends of the chain are pinned by the protein charges, the one-two molecules in the middle must be strongly frustrated orientationally. The result is the separation of the hydration shell into polarized domains with their low-temperature behavior reminiscent of the phenomenology of relaxor ferroelectrics close to the glass transition.\cite{DMjpcl:15} The key question relevant to this observation is that of the depth of propagation of these domains into the bulk, or, in other words, of the number of water molecules involved into the hydration layer with properties distinct from the bulk. Given that $\sim (300-500)$ molecules of water can be counted only in the first hydration layer of a typical globular protein, changes in the orientational structure of even the second hydration layer will produce an ensemble of water molecules  sufficiently large to be characterized as a separate mesophase ($\sim 1063$ water molecules within a 6 \AA\ shell around lysozyme\cite{DMjpcl:15}). Such a  shell will carry new properties requiring characterization. 

Absorption of radiation in the THz domain of frequencies has recently appeared as a novel technique to study hydration shells of proteins.\cite{Zhang:06,Xu:06,Heugen:06,Ebbinghaus:07,Born:09,Niehues:2011fk,Vinh:2011qf,Acbas:2014ci,Yamamoto:2016dv,Moix:2017ir} The absorption of light as the means of probing the solvation structure has a significant advantage compared to broad-band dielectric spectroscopy since the sensitivity of the response is improved by a factor roughly equal to the dielectric constant for the same magnitude of the liquid polarization or by a factor of dielectric constant squared for the same magnitude of the external electric field.\cite{DMpre:10}  The difference is due to the transverse geometry of light absorption compared to the longitudinal geometry of the dielectric experiment. 

Further advantage of the THz domain spectroscopy over absorption and dielectric spectroscopy at lower frequencies is that the response of the protein dipole (relaxation time $\sim 9$ ns for lysozyme\cite{DMjcp2:14}) is dynamically frozen at the THz frequency. While there are a number of vibrational modes in the THz region belonging to the vibrational density of states of a protein,\cite{Yu:2003hq,Lerbret:2012gz} they do not project on the protein dipole to produce sufficient THz absorption.\cite{DMjcp1:12} The protein is therefore mostly transparent in the THz domain and main absorption comes from the water component of the solution. The challenge is how to separate the hydration shell from the dominant bulk absorption. 

The analysis of the experimental data in terms of simplistic mixing model,\cite{Zhang:06,Heugen:06,Ebbinghaus:07,Luong:2011cn,Sushko:2015cv} disregarding non-additivity of the interfacial polarization (see discussion after Eq.\ \eqref{eq:3} below), suggested the possibility of an extended hydration shell around a protein, with absorption distinct from the bulk.\cite{Ebbinghaus:07,Luong:2011cn} It was later shown that the change in the absorption of solutions relative to bulk water can be related to the cross-correlation between the protein dipole and the dipoles of water molecules in the hydration shell.\cite{DMjcp1:12} Specifically, the alteration of the dipolar susceptibility of the solution relative to the bulk is directly proportional to the solute-solvent dipolar correlation function $\chi_{0s}(\omega)$
\begin{equation}
\Delta \chi(\omega)  \propto \chi_{0s}(\omega)
\label{eq}
\end{equation}
(see Eq.\ \eqref{eq:8} below for a complete form). When the cross-correlation $\chi_{0s}(\omega)$ is calculated for hydration shells of varying thickness, its length of saturation to the bulk value turns out to be in the range of 20--40 \AA\cite{DMjcp1:12,DMjcp3:12} from the protein surface. Therefore, deviations from the ideal behavior with increasing solute concentration\cite{Ebbinghaus:07} are likely related to changes of $\chi_{0s}$ from its infinite-dilution value.             

\begin{figure}
\includegraphics*[width=5cm]{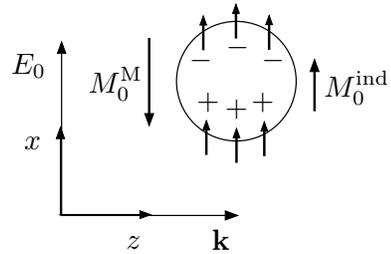}
\caption{Schematic representation of the polarization of the solute-solvent interface by the uniform field of radiation propagating with the wave-vector $\mathbf{k}$ along the $z$-axis and polarized along the $x$-axis. We assume the slab geometry, with $\mathbf{E}_0$ parallel to the plane of the slab. The short arrows, aligned along the field, indicate dipoles of water polarized by the field, which create surface charge density with the positive and negative lobes at the opposite sides of the cavity. The surface charge density obtained by solving the standard dielectric boundary value problem\cite{Jackson:99} results in the interface dipole $M_0^\text{M}$ given by Eq.\ \eqref{eq:1}. The dipole $M_0^\text{ind}$, induced at the solute by the external field, is the sum of the electronic and nuclear components, Eq.\ \eqref{eq:0}. The permanent solute dipole is strongly affected by the radiation frequency and is effectively zero for a protein in the electric field of THz radiation.  
}
\label{fig:1}  
\end{figure}

As expected from the general arguments presented above, simulations have shown that  orientational correlations of water dipoles in the hydration layer are significantly different from those in a homogeneous dielectric. The major question, still posed to the field, is whether these differences can be reliably extracted from absorption of THz radiation with its wavelength far exceeding the length-scale of any structural correlations in the solution. Since long-wavelength radiation does not provide spatial resolution, theory is required to interpret the data. 

The first question to ask is what would the standard electrostatics predict for the absorption coefficient of solutions. This problem can in fact be addressed by  Maxwell electrostatics of dielectrics, by which we mean the application of the Maxwell boundary conditions to the solution of the Poisson equation.\cite{Jackson:99} The problem at hand is illustrated in Fig.\ \ref{fig:1}. The polarization of the interface of a spherical void approximating the hydrated protein integrates to an effective dipole assigned to the void, which is anti-parallel to the external field
\begin{equation}
M_0^\text{M} = - \Omega_0 P \frac{3\epsilon_s}{2\epsilon_s+1} .
\label{eq:1}  
\end{equation}
Here, $P$ is the polarization of the bulk caused by the uniform field of the electromagnetic wave, $\Omega_0$ is the volume of the spherical solute and $\epsilon_s$ is the dielectric constant of water; the superscript ``M'' specifies the solution of the Maxwell boundary value problem. In contrast to this prediction, simulations\cite{DMjcp2:11,DMjcp1:12} and absorption measurements for hydrated lysozyme\cite{Novelli:2017bs} and for some amino acids\cite{Niehues:2011fk} have suggested that the polarization of the interface can significantly deviate from the Maxwell scenario. We indeed show below that the so-called Lorentz void,\cite{DMepl:08} instead of the Maxwell void (Eq.\ \eqref{eq:1}), provides a better representation of the collective interfacial polarization. However, this conclusion cannot be reached without taking into account the induced dipole of the protein itself. 

The negative projection of the interface Maxwell dipole is counterbalanced by a positive induced solute dipole $M_0^\text{ind}$. The latter is composed of the instantaneous (on the THz time-scale) electronic induced dipole $M_0^e$ and the average permanent dipole  $\langle M_0\rangle_E$ established along the field within an ensemble of proteins in solution
\begin{equation}
M_0^\text{ind} = M_0^e + \langle M_0\rangle_E .
\label{eq:0}  
\end{equation}
The permanent dipole of course depends on the frequency of the signal and is expected to be dynamically frozen in the THz domain. The average over an ensemble of randomly oriented protein permanent dipoles then produce a negligible net value of $\langle M_0\rangle_E$. Nevertheless, dynamical freezing depends on the radiation frequency and should not be blindly assumed. The induced dipole cannot be generally neglected, as we indeed show below. Given the mutual compensation between the anticipated negative interface dipole and the solute induced dipoles, a quantitative theory is required to incorporate both components.  Here, we construct such a theoretical description focused on experimental applications and based on previous theoretical advances.\cite{DMpre:10,DMjcp1:12} The frequency dependence of the permanent protein dipole induced by the external field is supplied by molecular dynamics (MD) simulations. These results confirm that the protein permanent dipole is dynamically frozen in the THz frequency window, allowing one to neglect $\langle M_0\rangle_E$ in Eq.\ \eqref{eq:0}. The main compensation therefore happens between the positive induced electronic dipole and the negative interface dipole. We compare the results of our analysis to recent THz absorption data reported for lysozyme solutions.\cite{Novelli:2017bs}

\section{Formalism}
\label{sec:2}
The absorption coefficient for electromagnetic radiation at frequency $\omega$ is fully defined in terms of the real and imaginary parts of the dielectric constant of the solution $\epsilon(\omega)=\epsilon'(\omega)+i\epsilon''(\omega)$ 
\begin{equation}
\alpha_\text{abs}(\omega) = \frac{\omega}{c} \frac{\epsilon''(\omega)}{\sqrt{\epsilon'(\omega)}} .
\label{eq:2}
\end{equation}
The problem is therefore reduced to calculating $\epsilon(\omega)$ in terms of the complex-valued dielectric constant of the solvent $\epsilon_s(\omega)$ (water in our case) and the properties of the solute (protein). The general formalism for $\epsilon(\omega)$ of an ideal solution was developed in Refs.\ \onlinecite{DMpre:10,DMjcp1:12}. Here we summarize the main steps of this theory relevant to our simulations and comparison to experiment.
 
We first consider a fictitious solute which does not produce any dipolar response and thus can be approximated as a void in a polar liquid. This approximation is relevant to many THz experiments when the solute dipole is too slow and is dynamically frozen. Computer simulations presented below specifically address the question of whether this approximation is applicable in the case of hydrated lysozyme. Independently of the outcome, this limit is the simplest conceptual point of departure for our discussion. 

\subsection{Voids in water}

A void removes polar material from its volume thus reducing the overall dipole moment of the liquid from its homogeneous magnitude $M^\text{liq}_x=PV$ to the value $M_x^\text{liq}- N_0 \Omega_0 P$. Here, we have adopted the geometry of an absorption experiment sketched in Fig.\ \ref{fig:1} in which light is propagated with the wave-vector along the $z$-axis of the laboratory frame and its electric field polarized along the $x$-axis. Further, the solution contains $N_0$ solutes, each having the volume $\Omega_0$, $\eta_0=(N_0\Omega_0)/V$ is the volume fraction. 

Removing polarized medium from the void's volume is not sufficient to calculate the polarization of the solution. Any discontinuous interface inside a dielectric is polarized by the field, and there will be an additional interfacial polarization, which requires solving the dielectric boundary value problem.\cite{Jackson:99} In the case of a spherical void, the  transverse geometry of the absorption experiment leads to the following relation\cite{DMpre:10}
\begin{equation}
M_x = P(V - N_0\Omega_0)  - N_0\Omega_0 P \frac{\epsilon_s-1}{2\epsilon_s+1} .
\label{eq:3}
\end{equation}
Here, the last summand accounts for an additional polarization of the interface not accounted for by the removal of the polarized liquid. 

The first term in Eq.\ \eqref{eq:3} projects on the reduction of absorption  by removing water from the volume of the protein. This is the water term in the two-component analysis often used in the literature.\cite{Zhang:06,Heugen:06,Ebbinghaus:07,Luong:2011cn,Sushko:2015cv} In this approach, the absorption coefficient of the solution is represented as the volume-fraction weighted sum of the protein (p) and water (w) components: $\alpha_\text{abs}=(1-\eta_0)\alpha_w+\eta_0\alpha_p$. This approximation can produce negative protein absorption when applied to analyzing experimental data.\cite{Vinh:2011qf} This seemingly astonishing result is in fact trivial when compared to Eq.\ \eqref{eq:3}: the additional negative contribution, which is assigned to absorption by the protein in the two-component analysis, is caused by polarization of the void in addition to the removal of water (the second summand in Eq.\ \eqref{eq:3}). 

The two-component analysis often leads to seemingly reasonable results, but this comes not from the soundness of this approximation, but instead from the combination of the breakdown of the Maxwell dielectric picture for the heterogeneous protein-water interface and a potential compensating effect of the dipole induced at the solute ($M_0^\text{ind}$ in Eq.\ \eqref{eq:0} and in Fig.\ \ref{fig:1}, also see below).\cite{DMjcp1:12} Deviations between the two-component model and observations are often assigned to the third component, the hydration water. Given that such deviations, according to Eq.\ \eqref{eq:3}, are predicted even within the standard dielectric picture, any conclusions based on such an analysis are unfounded. Numerically, the additional polarization of the interface  [second summand in Eq.\ \eqref{eq:3}] is a significant effect, amounting to about a half, at $\epsilon_s\gg1$, of the polarization loss from the direct expulsion of water from the solute volume.         

The appearance of the dielectric constant of bulk water $\epsilon_s$ in Eq.\ \eqref{eq:3} is a consequence of using the standard dielectric boundary conditions at the discontinuous surface separating the void from the polar liquid. Hydration water surrounding proteins is strongly altered compared to the bulk and it is a priori not clear if the standard dielectric boundary conditions apply. In fact, we have shown before\cite{DMjcp1:12} and demonstrate more conclusively here that polar response of the protein-water interface is qualitatively different from that of the standard dielectric interface. In order to account for these differences, we introduced the parameter $\alpha(\omega)$, which allows us to re-write  Eq.\ \eqref{eq:3} in the following form
\begin{equation}
\begin{split}
M_x(\omega) &= P(\omega)(V - N_0\Omega_0 ) \\
&- N_0\Omega_0 \alpha(\omega) P(\omega) \frac{\epsilon_s(\omega)-1}{2\epsilon_s(\omega)+1} .
\end{split}
\label{eq:4} 
\end{equation}
Here, we have indicated the dependence of all components on the radiation frequency $\omega$. It has to be stressed that, similarly to the dielectric constant $\epsilon_s(\omega)$, the parameter $\alpha(\omega)=\alpha'(\omega)+i\alpha''(\omega)$ is a complex-valued function, as will be clear from the following discussion.   

From Eq.\ \eqref{eq:4}, one can directly calculate the change in the susceptibility of the solution $4\pi\chi(\omega)=\epsilon(\omega)-1$ relative to the susceptibility of the bulk liquid $4\pi\chi_s(\omega)=\epsilon_s(\omega)-1$
\begin{equation}
\frac{\Delta\chi_\text{v}(\omega)}{\chi_s(\omega)} = - \eta_0\left[1+\alpha(\omega) \frac{\epsilon_s(\omega)-1}{2\epsilon_s(\omega)+1} \right] ,
\label{eq:5}
\end{equation}
where we used the subscript ``v'' to stress that this susceptibility change is due to the ideal solution of voids. In the case of the Maxwell dielectric, $\alpha(\omega)=1$ and the solution susceptibility decreases linearly with the volume fraction $\eta_0$ 
\begin{equation}
\frac{\Delta\chi_\text{v}(\omega)}{\chi_s(\omega)} = - \eta_0 \frac{3\epsilon_s(\omega)}{2\epsilon_s(\omega)+1} .
\label{eq:6} 
\end{equation}
The slope of the linear scaling with $\eta_0$ is equal to the the product of $\epsilon_s$ with the cavity field susceptibility of dielectric theories\cite{Boettcher:73,Frohlich} 
\begin{equation}
\chi_c^\text{M}(\omega) = \frac{3}{2\epsilon_s(\omega)+1} 
\label{eq:6-1}
\end{equation}
The susceptibility $\chi_c=E_c/E_0$ is the ratio of the field inside the void (cavity) $E_c$ to the uniform external $E_0$ (see below), the superscript ``M'' indicates the Maxwell solution for this property.\cite{Jackson:99,DMepl:08}  
 
Equation \eqref{eq:5} does not accomplish much unless a connection between $\alpha(\omega)$ and parameters accessible by either simulations and/or experiment can be established. It was suggested\cite{DMjcp1:12} that the effective dipole established at the surface of a spherical void polarized by the uniform external field can be alternatively calculated from the response to a dipole placed inside the void. The connection is provided by the relation
\begin{equation}
 1+\alpha(\omega) \frac{\epsilon_s(\omega)-1}{2\epsilon_s(\omega)+1} = -
 \frac{3\epsilon_s(\omega)}{2(\epsilon_s(\omega)-1)} \, \frac{\chi_{0s}(\omega)}{\chi_{00}(\omega)} .
 \label{eq:7} 
\end{equation}
Equation \eqref{eq:5} is therefore re-written in the following form
\begin{equation}
  4\pi\Delta\chi_\text{v}(\omega) = \tfrac{3}{2} \eta_0 \epsilon_s(\omega) \frac{\chi_{0s}(\omega)}{\chi_{00}(\omega)} . 
  \label{eq:8} 
\end{equation}

Equation \eqref{eq:8} provides a significant advantage for the theory-experiment connection since it directly yields the THz slope ($\alpha_\text{abs}$ vs $\eta_0$), instead of absorption at a given concentration as calculated from simulations of the total dipole of the mixture.\cite{Heyden:08,Moix:2017ir}  The required input is two susceptibility functions, $\chi_{00}(\omega)$ and $\chi_{0s}(\omega)$. They are accessible by computer simulations in the limit of a single solute in the simulation cell, thus bypassing the difficulties of simulating mixtures. The susceptibilities $\chi_{00}(\omega)$ and $\chi_{0s}(\omega)$ arise from, correspondingly, the ``self'', $\propto\langle \delta \mathbf{M}_0(t)\delta \mathbf{M}_0(0)\rangle$, and ``cross'', $\propto\langle \delta \mathbf{M}_s(t)\cdot\delta \mathbf{M}_0(0)\rangle$, time correlation functions of the fluctuating dipole moments of the solute, $\mathbf{M}_0$, and solvent, $\mathbf{M}_s$. Being linear response functions, $\chi_{00}(\omega)$ and $\chi_{0s}(\omega)$  represent the permanent dipole moment induced by the external electric field at the solute (subscript ``00'') and the dipole moment caused in the hydration layer by the solute dipole (subscript ``0s''). 

The frequency-dependent susceptibilities are Laplace-Fourier $\omega$-transforms of the corresponding time correlation functions appearing in the dynamic linear response theory.\cite{Kubo:66}  One can define the normalized time correlation functions
\begin{equation}
S_{0a}(t)=\left[\langle \delta \mathbf{M}_0\cdot\delta\mathbf{M}_a\rangle \right]^{-1}  \langle \delta \mathbf{M}_0(t)\cdot\delta\mathbf{M}_a(0)\rangle,
\label{eq:9}
\end{equation}
where $\delta \mathbf{M}_a(t) = \mathbf{M}_a(t) - \langle\mathbf{M}_a\rangle$ and $a=0,s$. These functions are Laplace-Fourier transformed to yield the functions $\tilde S_{0a}(\omega)$, which then enter the corresponding linear response functions
\begin{equation}
\chi_{0a}(\omega)= (\beta/3\Omega_0)\langle \delta \mathbf{M}_0\cdot\delta\mathbf{M}_a\rangle \left[1 + i\omega \tilde S_{0a}(\omega) \right] . 
\label{eq:10} 
\end{equation}
The fitting procedure using a multi-exponential decay with amplitudes $A_i$ ($\sum_iA_i=1$) and relaxation times $\tau_i$ leads to
\begin{equation}
\tilde S_{0a}(\omega) = \sum_i  \frac{A_i\tau_i}{1-i\omega\tau_i}  
\label{eq:11}
\end{equation}
and 
\begin{equation}
\chi_{0a}(\omega) \propto \sum_i A_i \left[1-i\omega \tau_i \right]^{-1} .  
\label{eq:12} 
\end{equation}

The cross correlations $\propto\langle \delta \mathbf{M}_s(t)\cdot\delta \mathbf{M}_0(0)\rangle$ are typically poorly converged and are a challenge to calculate from simulations. In anticipation of the final results used to calculate the absorption coefficients, we note that the response function
\begin{equation}
 \chi_0^d(\omega)=\chi_{00}(\omega)+\chi_{0s}(\omega)
 \label{eq:12-1}
\end{equation}
enters the final expressions. This function is based on the time correlation function $\propto\langle \delta \mathbf{M}_0(t)\cdot \delta \mathbf{M}(0)\rangle$ between the solute dipole and the entire dipole moment of the sample $\mathbf{M}$, which is typically a faster converging property. 

\begin{figure}
\includegraphics*[width=4cm]{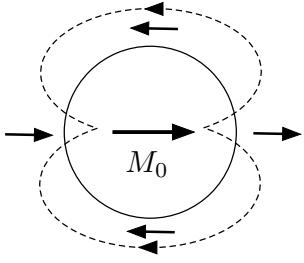}
\caption{Schematic representation of the solvent dipoles oriented along the electric field lines (dashed) of the solute dipole $M_0$. The pole dipoles are aligned parallel to the solute dipole, and the equatorial dipoles are aligned antiparallel to the solute dipole. Since there are more equatorial than pole dipoles, the overall correlation function $\chi_{0s}(\omega)$ is typically negative.    
}
\label{fig:2}  
\end{figure}
 
Since $\chi_{00}$ is always positive, Eq.\ \eqref{eq:8} clearly shows that the sign of slope of $\Delta \chi_\text{v}$ vs $\eta_0$ is determined by the sign of the cross, solute-solvent dipolar correlations. The solute dipole and the dipoles of solvent molecules are expected to be anti-correlated, thus resulting in a usually observed\cite{Zhang:06,Xu:06,Novelli:2017bs} negative THz slope. The reason for such negative cross correlations is illustrated in Fig.\ \ref{fig:2}, which shows the dipoles of the solvent arranged around a solute dipole. While axial solvent dipoles tend to orient in parallel, the equatorial dipoles tend to orient anti-parallel. Since there are always more equatorial dipoles that axial dipoles, the overall cross-correlation is negative. This picture is, however, based on the assumption that interfacial dipoles can freely change their orientations along the field lines of the solute dipole. Hydration shells of proteins are densely packed\cite{Gerstein:1996tg,Svergun:98} and orientationally frustrated,\cite{DMjpcl2:12,DMjpcl:15} thus potentially affecting the ability of hydration waters to polarize along the field of the dipole. While most solutes still show a negative slope of $\Delta \chi_\text{v}$ vs $\eta_0$, solutions of some amino acids (glycine and serine\cite{Niehues:2011fk}) produce a positive THz slope. The latter result might be, at least partially, related to the compensating positive dipole moment induced at the solute (Eq.\ \eqref{eq:0}), which we consider below.     

\begin{figure}
\includegraphics*[clip=true,trim= 0cm 1cm 0cm 0cm,width=8cm]{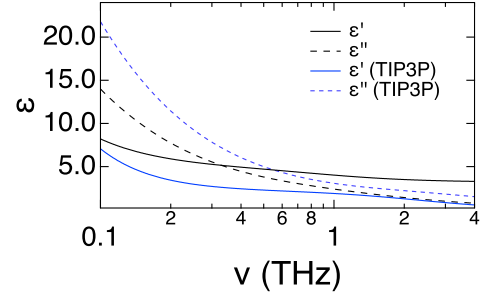}
\caption{Real, $\epsilon'(\nu)$, and imaginary, $\epsilon''(\nu)$, dielectric constants of water\cite{Vinh:2015hf} and of TIP3P force-field model of water in the THz range of frequencies. MD simulations of TIP3P water at $300$ K performed for this study were fitted with two Debye processes with dielectric increments $\Delta \epsilon_1=101.0$ and $\Delta \epsilon_2=2.2$ and relaxation times $\tau_1=7.05$ ps and $\tau_2=0.07$ ps.
}
\label{fig:3}  
\end{figure}

In the analysis below, we use the input for the frequency-dependent dielectric constant of water from experiment\cite{Vinh:2015hf} and model the electronic induced dipole of the protein by its refractive index. The force fields typically used in MD simulations miss polarizability of at least one of these two components. It is not a priori clear if the force fields typically used in condensed phase simulations are suitable for modeling absorption. To illustrate this point, Fig.\ \ref{fig:3} compares the real and imaginary parts of $\epsilon_s(\nu)$ ($\omega=2\pi\nu$) for TIP3P water (dashed lines) with experiment\cite{Vinh:2015hf} (solid lines).  The differences between experiment and TIP3P water are significant. In order to establish whether these differences affect absorption of radiation, one needs a formalism operating in terms of solvent parameters accessible by both experiment and simulations.  Equation \eqref{eq:8} presents this opportunity, and we discuss this analysis below. It turns out that the use of the dielectric constant of TIP3P water significantly deteriorates the agreement between calculations and experiment within the analytical (Lorentz void) framework, but does not strongly affect the results directly obtained from MD simulations.      

\subsection{Solute dipole} 
The solute dipole can include both the permanent and electronic components. As is typically done in mean-field theories of polarizable dielectrics,\cite{Wertheim:1979ty,SPH:81} it is convenient to combine them in one polar density parameter\cite{DMjcp1:12}
\begin{equation}
y_0(\omega) =  (4\pi/3) \chi_{00}(\omega) + y_e,  
\label{eq:13}
\end{equation}
where $y_e=(4\pi/3)(\alpha_0/\Omega_0)$ and $\alpha_0$ is the electronic polarizability of the solute. Clausius-Mossotti equation can be used to estimate $y_e$
\begin{equation}
  y_e=\frac{1}{\eta_0^p}\frac{n_p^2-1}{n_p^2+2} .
  \label{eq:13-1}
\end{equation}
Here, $n_p$ is the refractive index of the protein and $\eta_0^p$ is the packing fraction of proteins in the powder used to measure $n_p$. We will assume $\eta_0^p\simeq 1$ in our estimates below, which is also the assumption made by Onsager for molecular polarizability.\cite{Onsager:36} Corrections can be introduced if this parameter is known independently.   

The frequency-dependent dipolar response of the solute adds to the solution susceptibility, which is now defined as
\begin{equation}
\Delta \chi = \Delta \chi_\text{v} + \Delta \chi_0  . 
\label{eq:14} 
\end{equation}
Here, the first summand, due to the void, is given by Eq.\ \eqref{eq:8}. The second terms, the change in the susceptibility due to dipolar response of the solutes, is given by the following equation\cite{DMjcp1:12} 
\begin{equation}
  \frac{4\pi\Delta\chi_0}{\eta_0y_0}  =
3\epsilon_s\chi_c  + J_0(\eta_0)\left[ 3\epsilon_s \chi_c - \epsilon_s - 2  \right],
\label{eq:15}  
\end{equation}
where here and in Eq.\ \eqref{eq:14} the dependence on frequency has been dropped for brevity. The frequency-dependent response function $\chi_c(\omega)$ is
\begin{equation}
\chi_c(\omega)=\frac{\chi_0^d(\omega)}{\chi_{00}(\omega)} = 1 +  \frac{\chi_{0s}(\omega)}{\chi_{00}(\omega)} ,
\label{eq:15-1}
\end{equation}
where $\chi_0^d(\omega)$ given by Eq.\ \eqref{eq:12-1} is the total susceptibility of the solute dipole in response to the uniform polarizing external field.\cite{Stern:2003cn,DMcpl:11}  The physical meaning of $\chi_0^d/\chi_{00}$ is that it is equal\cite{DMjcp1:12} to the ratio, $E_c/E_0$, of the cavity field $E_c$ experienced by the solute dipole from the polarized solvent to the field of the external charges $E_0$ (the result of Maxwell electrostatics is given by Eq.\ \eqref{eq:6-1}). Although a direct measurement of the cavity field is hardly possible, it enters a number of spectroscopic observables.\cite{Toptygin:2003ly} Correspondingly, the response function $\chi_c(\omega)$ represents the ratio of the cavity field to the external field at a given frequency.      

The first summand in Eq.\ \eqref{eq:15} is the infinite-dilution contribution of the solute dipole to the solution susceptibility. The second summand, which includes the function $J_0(\eta_0)$, represents the non-ideal effect of polarizing the void by the permanent dipoles of other solutes in the solution. Therefore, $J_0(0)=0$, and it is given for an arbitrary $\eta_0$ by an equation involving the density structure factor of the solutes, $S_0(k)= N_0^{-1}\sum_{i,j}^{N_0}\exp[i\mathbf{k}\cdot \mathbf{r}_{ij}]$, where $\mathbf{r}_{ij}$ is the vector connecting two solutes in the solution and $\mathbf{k}$ is the wavevector. The density structure factor determines mutual correlations of the positions of the solutes and $J_0(\eta_0)$ is given as\cite{DMpre:10}  
\begin{equation}
J_0(\eta_0)= \frac{3}{\pi} \int_0^{\infty} [j_1(x/2)]^2 \left[S_0(x)-1\right] dx .
\label{eq:16}  
\end{equation}
Here, $j_1(x)$ is the spherical Bessel function of the first order\cite{Abramowitz:72} and integration is performed over the dimensionless variable $x=k\sigma_0$, where $\sigma_0$ is the solute diameter. While $S_0(k)$ is experimentally available from  small-angle scattering, we estimated it here based on Percus-Yevick solution for a fluid of hard spheres.\cite{Hansen:03} These estimates (see $J_0(\eta_0)$ tabulated in Ref.\ \onlinecite{DMpre:10}) show that $J_0(\eta_0)$ can be dropped from the final equations at the low volume fraction $\eta_0<0.1$. When  $J_0=0$,  one can apply Maxwell electrostatics ($\alpha=1$) as a consistency check of Eq.\ \eqref{eq:15}. One then gets Eq.\ \eqref{eq:6-1} for $\chi_{0}^d/\chi_{00}$  and for the static ($\omega=0$) response 
\begin{equation}
\Delta \chi_0 = N_0 \frac{\beta \langle \delta\mathbf{M}_0^2\rangle}{3 V} \, \frac{3\epsilon_s}{2\epsilon_s+1} .
\label{eq:16}  
\end{equation}
This is the standard linear-response result\cite{Boettcher:73} for an ideal solution of impurities experiencing the cavity field $E_c=3\epsilon_sE/(2\epsilon_s+1)$ from the surrounding dielectric (cf.\ to Eq.\ \eqref{eq:6-1}, $E$ is the Maxwell field in the bulk).

\section{Simulation results}

All-atom MD simulations were performed using the initial crystallographic structure of lysozyme (PDB entry 1AIK) solvated with TIP3P water. The simulation cell consisted of a total of 87050 atoms, with 28361 water molecules. The total charge of the lysozyme protein was $-7$ e.  The constant pressure temperature equilibration simulations were done using the Langevin temperature-pressure control with the damping coefficient of 5 ps$^{-1}$, a piston pressure of 1 atm, a piston decay time of 50 fs, a piston oscillation period of 200 fs, and at temperature of 300 K. A cutoff radius of 12  ̊\AA\ and full electrostatics using the particle mesh Ewald technique at every simulations step were employed. NAMD 2.10\cite{Phillips:2005qv} with the CHARMM27 force field was used to produce the MD trajectories. An initial optimization of the simulation cell was performed by conjugate gradient minimization for 2000 steps, followed by a 5 ns NPT equilibration simulation. Proceeding from equilibration run, production NVT simulations were carried out for 10 ns.

\begin{figure}
\includegraphics*[clip=true,trim= 0cm 0cm 0cm 0cm,width=8cm]{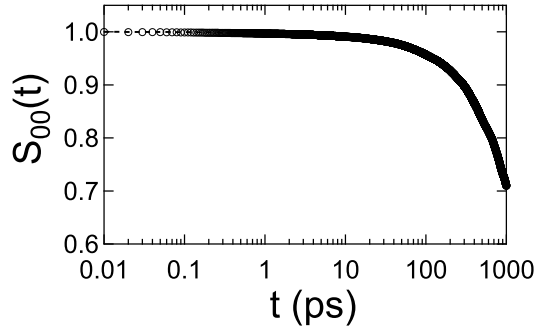}
\caption{Time correlation function $S_{00}(t)$ [Eq.\ \eqref{eq:9}] calculated from MD simulations (points) and fitted by three decaying exponents (dashed line): $S_{00}(t)=\sum_{i=1}^3A_i\exp[-t/\tau_i]$: $A_1=0.004$, $A_2=0.006$, $A_3=0.990$, $\tau_1=0.77$ ps, $\tau_2=24.6$ ps, $\tau_3=2988$ ps.    
}
\label{fig:4}  
\end{figure}

The main reason for a relatively short length of the simulation trajectory is that the simulations (both the equilibration NPT and the production NVT) were performed with a relatively short timestep of 0.5 fs and flexible hydrogen bonds of the protein. The trajectory was saved every 10 fs. This focus on the short-time dynamics has allowed us to calculate the time correlation functions with a 10 fs time resolution required for modeling the THz response. Exponential fits of the correlation function were done on the time window of 1 ns. Those were Laplace-Fourier transformed to obtain Eq.\ \eqref{eq:11} for the correlation functions and Eq.\ \eqref{eq:12} for the response functions.

\section{Comparison to experiment}
A significant result of our present and previous\cite{DMjcp1:12} computer simulations is that $(4\pi/3)\chi_{00}$ can be neglected in Eq.\ \eqref{eq:13} relative to $y_e$ in the THz domain. The nuclear motions of the protein do not produce sufficient fluctuations of its dipole and protein  absorption is negligible. Only the electronic dipole induced at the protein makes a non-negligible contribution. Note that, in contrast with the previous simulations,\cite{DMjcp1:12} the current setup allowed vibrations of protons in the protein, along with a significantly smaller integration step and more frequent savings. Nevertheless, $S_{00}(t)$ is nearly flat at $t\simeq 0.1-1$ ps (Fig.\ \ref{fig:4}) and, correspondingly, $\chi_{00}(\omega)$ is very small for $\nu = \omega/(2\pi)$ in the THz domain of frequencies.   

With this simplification and after dropping the term proportional to $J_0$ from Eq.\ \eqref{eq:15},  one arrives at a simple relation
\begin{equation}
4\pi\Delta \chi(\omega) = -\tfrac{3}{2}\eta_0\epsilon_s(\omega)\left[ 1 - \chi_c(\omega)\frac{3n_p^2}{n_p^2+2}\right],
\label{eq:17}  
\end{equation}
where the Clausius-Mossotti relation between the protein refractive index $n_p$ and $y_e$ was applied [$\eta_0^p=1$ in Eq.\ \eqref{eq:13-1}]. In the limit of the Maxwell electrostatics (superscript ``M'') one obtains
\begin{equation}
4\pi\Delta \chi^\text{M}(\omega) = -\tfrac{3}{2}\eta_0\epsilon_s(\omega)\left[ 1 - \frac{9n_p^2}{(2\epsilon_s(\omega)+1)(n_p^2+2)} \right] . 
\label{eq:18}  
\end{equation}
At $n_p=1$, this relation converts to Eq.\ \eqref{eq:6} for the susceptibility of a Maxwell void.

\begin{figure}
\includegraphics*[clip=true,trim= 0cm 1cm 0cm 0cm,width=8cm]{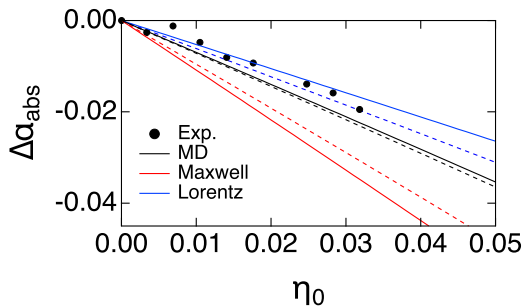}
\caption{Relative change of the absorption coefficient of the solution $\Delta\bar\alpha_\text{abs}(\omega)$ [Eq.\ \eqref{eq:20}] vs the solute volume fraction $\eta_0$. Calculations are based on $\chi_c)\omega)$ from MD (black, Eq. \eqref{eq:17}), and analytical equations based on Maxwell (red, Eq.\ \eqref{eq:18}) and Lorentz (blue, Eq.\ \eqref{eq:19}) routes. The points are experimental data\cite{Novelli:2017bs} and the calculations are made at $\nu=0.65$ THz. The solid lines refer to $n_p=1.55$\cite{Cervelle:1974kt} and the dashed lines correspond to the assumption of a non-polarizable protein, $n_p=1$. Experimental\cite{Vinh:2015hf}  $\epsilon_s(\omega)$ is used in all calculations.  
}
\label{fig:5}  
\end{figure}

One can additionally consider the limit in which no surface charges, originating from introducing a dividing surface into the dielectric,\cite{Jackson:99} are created at the interface of a void (no ``$+$'' and ``$-$'' at the void's surface in Fig.\ \ref{fig:1}). This limit corresponds to $\alpha(\omega)=0$ in Eq.\ \eqref{eq:5} and the assumption that the only result of creating the void is the expulsion of the polarized water from the void's volume. Such a void is known in the theory of dielectrics as the virtual, or Lorentz, cavity.\cite{Boettcher:73} We will therefore dub this limit as the ``Lorentz scenario''.\cite{DMepl:08}   The ratio $\chi_0^d/\chi_{00}$ becomes equal to the Lorentz cavity field,\cite{DMjcp1:12} $E_c/E_0=(\epsilon_s+2)/(3\epsilon_s)$, and one obtains from Eq.\ \eqref{eq:17} (superscript ``L'' is for the Lorentz scenario)
\begin{equation}
4\pi\Delta \chi^\text{L}(\omega) = -\tfrac{3}{2}\eta_0\epsilon_s(\omega)\left[ 1 - \frac{\epsilon_s(\omega)+2}{\epsilon_s(\omega)}\frac{n_p^2}{n_p^2+2} \right] .
\label{eq:19}  
\end{equation}
Not surprisingly, there is no change to the solvent response when the virtual void is filled with the same material as the solvent and $\Delta \chi^\text{L}=0$ at $\epsilon_s=n_p^2$. Note that the Lorentz cavity field is used in deriving the Clausius-Mossotti equation, and that fact leads to the complete cancellation of the interface and electronically induced solute dipoles at $\epsilon_s=n_p^2$. This does not happen in the Maxwell case in Eq.\ \eqref{eq:18} since the cavity field of dielectric theories,\cite{Boettcher:73} instead of the Lorentz field, is used in the Maxwell electrostatics. 

In application to analyzing experimental results, the complex-valued dielectric constant of bulk water was taken from recent measurements\cite{Vinh:2015hf} covering the range of frequencies $5.9-1120$ GHz. The experimental data\cite{Novelli:2017bs} for the absorption coefficient of lysozyme were obtained by  averaging absorption over the interval of frequencies $0.38-0.92$ THz. There is therefore a sufficient frequency overlap between the data for bulk water\cite{Vinh:2015hf} and solutions.\cite{Novelli:2017bs} The refractive index of the protein, $n_p\simeq 1.55$, is from Ref.\ \onlinecite{Cervelle:1974kt}.    

Figure \ref{fig:5} presents the calculations of the change in the absorption coefficient of the solution with increasing the volume fraction of the solute according to Eq.\ \eqref{eq:2}. Specifically, we normalize the change  in absorption by the absorption of bulk water
\begin{equation}
\Delta\bar\alpha_\text{abs}(\omega) = \alpha_\text{abs}(\omega)/\alpha_\text{abs}^s(\omega) - 1, 
\label{eq:20}  
\end{equation}
where the absorption of water $\alpha_\text{abs}^s(\omega)$ is calculated from Eq.\ \eqref{eq:2} with the water dielectric constant $\epsilon_s(\omega)$ used in place of the solution dielectric constant $\epsilon(\omega)$. The solution dielectric constant is calculated as $\epsilon(\omega)=\epsilon_s(\omega)+4\pi\Delta\chi(\omega)$, with three scenarios presented by Eqs.\ \eqref{eq:17}--\eqref{eq:19} used  for $\Delta\chi(\omega)$.  

\begin{figure}
\includegraphics*[clip=true,trim= 0cm 1cm 0cm 0cm,width=8cm]{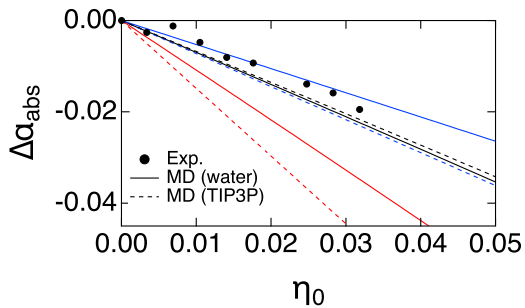}
\caption{Relative change of the absorption coefficient of the solution $\Delta\bar\alpha_\text{abs}(\omega)$ [Eq.\ \eqref{eq:20}] vs the solute volume fraction $\eta_0$. Calculations are based on $\chi_c(\omega)$ from MD (black, Eq. \eqref{eq:17}),  Maxwell (red, Eq.\ \eqref{eq:18}) and Lorentz (blue, Eq.\ \eqref{eq:19}) routes. The points are experimental data\cite{Novelli:2017bs} and the calculations are made at $\nu=0.65$ THz. The solid lines refer to $n_p=1.55$\cite{Cervelle:1974kt} and the experimental results for and the dielectric constant of water.\cite{Vinh:2015hf} The dashed lines refer to the same conditions, but $\epsilon_s(\omega)$ from MD simulations of TIP3P water (this study).  
}
\label{fig:6}  
\end{figure}

Both the Maxwell and Lorentz voids are sensitive to the choice of the protein refractive index. The dashed lines in Fig.\ \ref{fig:5} refer to the calculations neglecting the induced dipole of the protein ($M_0^\text{ind}=0$ in Eq.\ \eqref{eq:0}), which is achieved by putting $n_p=1$. There is, however, no physical reason to adopt this assumption and this limit is shown here as a mere warning that a ``good agreement'' with experiment can follow from unphysical approximations (the Lorentz scenario is in perfect agreement with experimental data in this case, blue dashed line). It is also clear that MD simulations of $\chi_c(\omega)$ provide a good account of the cavity field dynamics and are consistent with experiment. We additionally show in Fig.\ \ref{fig:6} the comparison of the calculations employing the experimental\cite{Vinh:2015hf} dielectric constant $\epsilon_s(\omega)$ (solid lines) and the same function obtained from MD simulations of TIP3P water (dashed lines). While there is little sensitivity to the choice of the water model in simulations, which access the dynamics of the cavity field directly from $\chi_c(\omega)$, the Maxwell and Lorentz equations for the cavity field are obviously sensitive to the choice of the frequency-dependent dielectric constant function. The agreement with experiment is much worse when $\epsilon_s(\omega)$ for TIP3P water is used Eqs.\ \eqref{eq:18} and \eqref{eq:19}. This is not surprising given large discrepancy in the dielectric properties between the experimental and TIP3P water shown in Fig.\ \ref{fig:3}.

\section{Conclusions} 
Absorption of radiation provides access to the orientational structure of the hydration shell in terms of the dipolar solute-solvent (cross) correlation function  [Eqs.\ \eqref{eq} and \eqref{eq:8}]. The structure of water can be disrupted to a different extent depending on the solute. The extent of the disruption and the resulting structure of the hydration shell affect the decay of the solute-solvent correlations into the bulk.\cite{DMjcp1:12} The decay length can be in the range of $20-40$ \AA.  The extended hydration shell anticipated in the past\cite{Ebbinghaus:07} in fact implies an extended cross-correlation.  Nuclear motions of the protein do not contribute to the signal and protein itself is effectively transparent in the THz domain.  Therefore, it is the protein-water interface that determines the change of THz absorption by solutions relative to the bulk.  

The analytical model\cite{DMpre:10,DMjcp1:12} for the absorption of radiation by solutions gives the slope of the absorption coefficient vs the solute volume fraction  in terms of the cavity response function $\chi_c(\omega)$ [Eq.\ \eqref{eq:17}], which is the ratio of the field inside the solute to the external field of radiation measured at the radiation frequency.  Simulations of a single lysozyme in solution performed here are capable of producing this function and lead to a good agreement with experiment\cite{Novelli:2017bs} (Fig.\ \ref{fig:5}). 

In the absence of direct simulations, two analytical limits, corresponding to the Maxwell and Lorentz voids, can be produced [Eqs.\ \eqref{eq:18} and \eqref{eq:19}]. We find, in agreement with previous calculations,\cite{DMjcp3:12} that the Maxwell void gives a poor representation of the data, while the Lorentz void gives a better account of experiment. One still has to realize that the appearance of the Lorentz limit for the void polarization is not consistent with the standard electrostatics, but is an emergent consequence of the orientational and density restructuring of the protein hydration layer compared to bulk water. It is therefore the deviation from the Maxwell limit that should be considered as the specific effect of the protein hydration shell. The Lorentz scenario is an effective-medium representation of the complex structure of the protein-water interface, which requires modification of the standard dielectric boundary-value problem.

\acknowledgments 
This research was supported by the National Science Foundation (CHE-1464810) and through XSEDE resources (TG-MCB080116N). 

\appendix


%

\end{document}